\documentclass[11pt]{article}

\usepackage[T1]{fontenc}
\usepackage[utf8]{inputenc}
\usepackage{lmodern}
\usepackage[margin=1in]{geometry}
\usepackage{graphicx}
\usepackage{booktabs}
\usepackage{amsmath}
\usepackage{array}
\usepackage[numbers,sort&compress]{natbib}
\usepackage{xcolor}
\usepackage[colorlinks=true,linkcolor=blue!50!black,citecolor=blue!50!black,urlcolor=blue!50!black]{hyperref}
\usepackage{caption}
\title{Comparing Experiences of Speech-Driven Game Modification\\
in Solo Play, Paired Play, and Paired Watching\\
Using the Fighting Game PATCH FIGHTER}

\author{Maya Grace Torii\thanks{Corresponding author: \href{mailto:toriparu@digitalnature.slis.tsukuba.ac.jp}{toriparu@digitalnature.slis.tsukuba.ac.jp}}\\
\normalsize University of Tsukuba\\
\normalsize Tsukuba, Japan
\and Takahito Murakami\\
\normalsize University of Tsukuba\\
\normalsize Tsukuba, Japan
\and Yoichi Ochiai\\
\normalsize University of Tsukuba\\
\normalsize Tsukuba, Japan}

\date{}

\begin{document}
\maketitle

\begin{abstract}
Players often have ideas about a game while playing it, but turning them into changes usually means stepping out of play to edit it.
We built PATCH FIGHTER, a two-player fighting game defined by 91 types of adjustable parameters.
A large language model (LLM) reads players' conversations during play and rewrites these parameters between games, so players can change the game just by talking.
We conducted a within-subjects experiment with 27 pairs (54 participants) comparing three conditions, \emph{solo play}, \emph{paired play}, and \emph{paired watching}.
Participants rated enjoyment after each game and answered a questionnaire on speech and changes after each condition.
Paired play was the most enjoyable condition.
In paired play, participants also felt more able to say what they thought about the game than in solo play.
In interviews, participants valued talking with the partner and checking changes through their own play.
Paired watching was less preferred because they could not try the changes themselves.
As our design cannot separate the enjoyment of editing from that of play, we outline a design that measures them separately and the evaluation of the resulting games as future work.
\end{abstract}

\noindent\textbf{Keywords:} generative AI, computational co-creativity, player-participatory game design, speech interfaces, within-subjects study

\section{Introduction}

The rules and content of a game are usually fixed by its makers before anyone plays.
Complaints and ideas that players have during play do not reach the game directly.
In a typical development process, player feedback is collected through playtesting, analyzed and prioritized by the development team, implemented in a new build, and then evaluated in further playtests.
Large language models (LLMs) can translate natural language into game content and changes to game settings~\cite{gallotta2024llmgames,sudhakaran2023mariogpt,gallotta2024llmaker,whitehead2025conversational}.
This could shorten the cycle from player feedback to a changed game.
However, even when a system changes the game in response to what players say, the players may not feel that the changes are their own design, whether individually or as a pair.
In computational co-creativity, how the input is formed and how the relationship to the generated output is understood can affect the sense of agency, ownership, and authorship~\cite{rezwana2023cofi,draxler2024ghostwriter}.

In our previous work, Lottery and Sprint~\cite{torii2023lottery} and LaS-Arcade~\cite{torii2026arcade}, non-experts designed games with generative AI.
Lottery and Sprint showed that teams including participants with no board-game design experience could generate and playtest board games through AutoGPT and the Design Sprint method.
However, making and playing remained separate stages.
LaS-Arcade then let a single player edit the settings of a Space Invaders--style game by voice while playing, shortening the play--edit--feedback cycle.
Yet in both studies, participants still told the AI explicitly what to change.

We frame the next step as a shift \textbf{from commands to discussion}, and \textbf{from a single speaker's utterances to requests that take shape in a conversation between two players}.
The two players do not give the AI editing commands in turn.
While watching or playing the game, they talk about what is fun, what feels off, what frustrates them, and how the game should be, agreeing, disagreeing, adding to, and rephrasing each other.
The system reads this natural conversation together with the play record, translates it into design decisions, and turns them into changes between games.

The underlying question of this research is \textbf{what kind of experience it is to change a game's settings through conversation while playing it}.
Having more chances to talk about the game, getting the desired changes, enjoying the editing itself, and enjoying the game after the change are possibly related but distinct experiences.
For example, a high enjoyment rating after a game may come from multiple factors, such as winning, talking with the partner, or the editing itself.
In this paper, we examine how the experience of changing the game differs depending on who one participates with and how.
Because our design cannot separate the enjoyment of editing from that of play, we interpret the ratings as reflecting the overall experience of playing while changing the game.

We pose the following two research questions.
\begin{description}
\item[RQ1 (social composition)] When participants play the game themselves, how do playing alone and playing as a pair differ in speech, game changes, per-game enjoyment, and the experience of having changed the game?
\item[RQ2 (form of engagement)] When participants talk as a pair, how do watching an AI-versus-AI match and playing against each other differ in speech, game changes, per-game enjoyment, and the experience of having changed the game?
\end{description}

To answer these questions, we designed a three-condition within-subjects study in which the same participants experienced solo play, paired play, and paired watching.
Comparing solo play with paired play is a social-composition contrast, which compares one and two players while both conditions involve playing.
Comparing paired play with paired watching is an engagement contrast, which compares playing and watching while both conditions involve two people talking.
We analyze the process in three stages, namely how participants talk, how the LLM turns their talk into changes, and how they experience the changed game.
In addition to the comparisons between conditions, we explore how the ratings relate to wins and losses, speech volume, and the number of changes, and use pair interviews to examine the background of these results.

The contributions of this paper are threefold.
(1) We built PATCH FIGHTER, which turns natural conversation between players into game changes.
(2) Through two pre-specified contrasts---solo versus paired play, and paired play versus paired watching---we provide local comparisons corresponding to the social composition of play and to the form of engagement when participating as a pair.
(3) Through pair interviews, we show why participants preferred paired play and avoided paired watching, namely talking with the partner and being able to check changes through their own play.

\section{Related Work}

\subsection{Runtime Generation and Editing of Game Content with Generative AI}
LLMs can play many roles in games, including player, design assistant, and content generator~\cite{gallotta2024llmgames}.
In procedural content generation (PCG), experience-driven PCG, which adapts content based on player experience, has been systematized~\cite{yannakakis2011edpcg}.
Since the advent of LLMs, researchers have reported methods that generate levels from natural language~\cite{sudhakaran2023mariogpt}, interfaces for designing levels with natural language alone~\cite{gallotta2024llmaker}, and attempts to operate procedural generators through conversation~\cite{whitehead2025conversational}.
Jennings et al. built a system that generates game behavior at runtime in response to player actions and examined its opportunities and challenges through interviews with game developers~\cite{jennings2024whatsthegame}. Studies that generate virtual environments by voice~\cite{liebers2024computer} and live game design that prototypes at the speed of play~\cite{vanrozen2025live} also aim to close the gap between making and playing.
Most of these assume a single user giving explicit instructions to the system.

\subsection{Improving Games through Agent Loops}
Game development with LLM agents has been proposed as frameworks in which multiple role-specialized agents divide the work~\cite{chen2023gamegpt}, and as conversational game engines that build a game through repeated dialogue with a user~\cite{hong2025chatge}.
Other studies run a loop of playing the generated game and revising it based on the results~\cite{huang2026play2code,wu2026rsigame}.
Play2Code repeatedly creates the next version from the findings of a GUI agent that plays the rendered screen~\cite{huang2026play2code}.
RSIGame combines a local loop that explores, diagnoses, and improves a game with a global loop that monitors quality over time and detects stagnation or regression~\cite{wu2026rsigame}.
This double loop is structurally similar to how PATCH FIGHTER rewrites settings at two levels, between games and between matches.
These works focus on automated loops but partly address human involvement.
Play2Code includes a condition in which a human plays instead of the GUI agent and returns revision reports, and RSIGame lets a human or a model play and write the policy for the next stage when improvement stalls.
In both, however, human input is given by one person in writing after play, and examining the effect of human involvement is not the aim.
Competitive games are also covered, but the opponent is an in-game AI, and evaluation focuses on scoring the generated games against rubrics.

\subsection{Conveying Intent in Natural Language and Forming Requests through Conversation}
Conveying intent to an LLM in natural language is difficult.
There is an abstraction gap between users' words and the representations a system requires~\cite{liu2023wants}, and non-AI experts try prompts opportunistically and overgeneralize from a few examples~\cite{zamfirescu2023johnny}.
Subramonyam et al.\ located the cause of this gap in translating intentions into prompts while it is still unclear what the task is, how to instruct, and what to expect~\cite{subramonyam2024gulf}.
In other words, the intent passed to a system is not necessarily complete from the start.
In conversation between people, by contrast, participants update their common ground while checking mutual understanding~\cite{clark1991grounding}, and deepen understanding while switching between the roles of explaining and questioning~\cite{miyake1986constructive}.
In creative work in which two people use generative AI together, vague ideas became concrete prompt wording as the pair talked~\cite{han2024teams}.
Agents that listen to a conversation between people without joining it and support it in the background have also been proposed, such as one that generates suggestions for the game master from the audio of a tabletop role-playing game~\cite{zhu2025overhearing}.
In these works, however, people write prompts themselves or the agent only assists the facilitator.
Changing the game itself from conversation between players has not been addressed.

\subsection{Playing Together and Watching}
Playing with others in the same place changes the game experience itself, and playing together in the same room has been reported to increase enjoyment~\cite{gajadhar2008shared}.
The design of the spectator experience~\cite{reeves2005spectator} and the pleasures of watching competitive games~\cite{cheung2011starcraft} have also been discussed, and in fighting games, audience participation has been reported to enhance the experience~\cite{paliyawan2024apg}.

\subsection{Positioning of This Work}
This work follows our previous studies on non-experts making games with generative AI.
In Lottery and Sprint, teams including participants with no board-game design experience generated board games with AutoGPT and the Design Sprint method and playtested them~\cite{torii2023lottery}.
Because making and playing were separate stages, each cycle of reading the generated rules, playing, and returning explicit feedback to the AI for regeneration took time.
LaS-Arcade let a single player edit the settings of a Space Invaders--style game by voice while playing, which shortened this cycle~\cite{torii2026arcade}.
These studies progressively moved game making by non-experts into play itself, but in both, participants told the AI explicitly what to change.
This work differs from these and the above studies in three ways.
First, the input that changes the game is not one person's explicit instruction but natural conversation in which players agree, disagree, and rephrase.
Second, this conversation is fed into the loop during play, game by game, and the changes are reflected in the next game.
Third, we compare forms of participation---solo play, paired play, and paired watching---through a user study rather than by scoring generated games.
In short, this work moves a single-player editing loop \textbf{from commands to discussion}, and \textbf{from a single speaker's utterances to requests formed in a conversation between two players}.

\section{PATCH FIGHTER}

\subsection{Architecture}
PATCH FIGHTER is a two-player fighting game whose entire behavior is determined by a single configuration object (\texttt{GameConfig}).
Physics, combat, visuals, sound, and each character's moves are represented as \textbf{91 modifiable axes}.
Each axis has a defined range (\texttt{CLAMP\_TABLE}).

\begin{figure}[t]
\centering
\includegraphics[width=\linewidth]{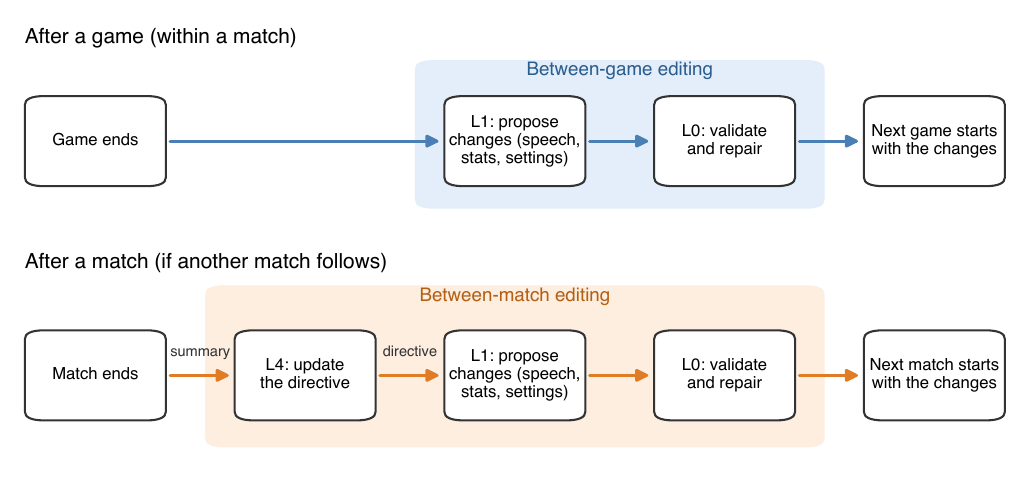}
\caption{When setting changes are executed. After a game within a match, L1 generates change proposals from the speech since the previous change, statistics from the last few games, and the current settings, and L0 validates and repairs them. When a match ends and another match follows, L4 updates the directive from a summary of that match and passes it to L1. Changes are applied at the start of the next game. After a match, the next match starts once L4, L1, and L0 have finished.}
\label{fig:system-timing}
\end{figure}

In this paper, a single win or loss is called a \textbf{game}, and a best-of-three set is called a \textbf{match}.
The settings are changed by three components that run at different times (Figure~\ref{fig:system-timing}).
Their labels follow the layer numbers of the PATCH FIGHTER system, and the other layers were not used in this study.
L1 is an LLM that runs after each game.
It reads the speech since the previous change, statistics from the last few games, and the current settings, and proposes changes to the settings.
L4 is an LLM that runs after each match.
It summarizes the match and updates a \textbf{directive}, a short natural-language policy for the game, which L1 receives as part of its instructions.
L0 is a rule-based validation step that checks every proposal from L1 before it is applied.
It clamps values to their ranges and rejects changes to nonexistent axes or of the wrong type.
The applied changes are shown on screen as a short announcement.

\subsection{Characters and Match Unit}
There are two characters, HADO and RUSH, with five initial moves.
The remaining six moves can be unlocked by L1 based on conversation and play.
A match is best of three and ends after two or three games.
Each time the condition changed, the settings were reset to the initial state, and the L4 directive was also local to each condition.

\section{Study}

\subsection{Design and Procedure}
Twenty-seven pairs of acquaintances (54 participants) took part in a within-subjects study in which everyone experienced the following three conditions.

\begin{itemize}
\item \textbf{Solo play.} A participant plays alone against an AI-controlled opponent. The two participants experience this condition individually.
\item \textbf{Paired play.} The two participants play against each other on the same computer.
\item \textbf{Paired watching.} The two participants watch an AI-versus-AI match together on the same computer.
\end{itemize}

Participants chose the CPU strength from ``weak,'' ``normal,'' and ``strong.''
It was not manipulated as an experimental factor.
The order of conditions was counterbalanced so that order effects would not be concentrated on any particular condition.

\begin{figure}[t]
\centering
\includegraphics[width=0.85\linewidth]{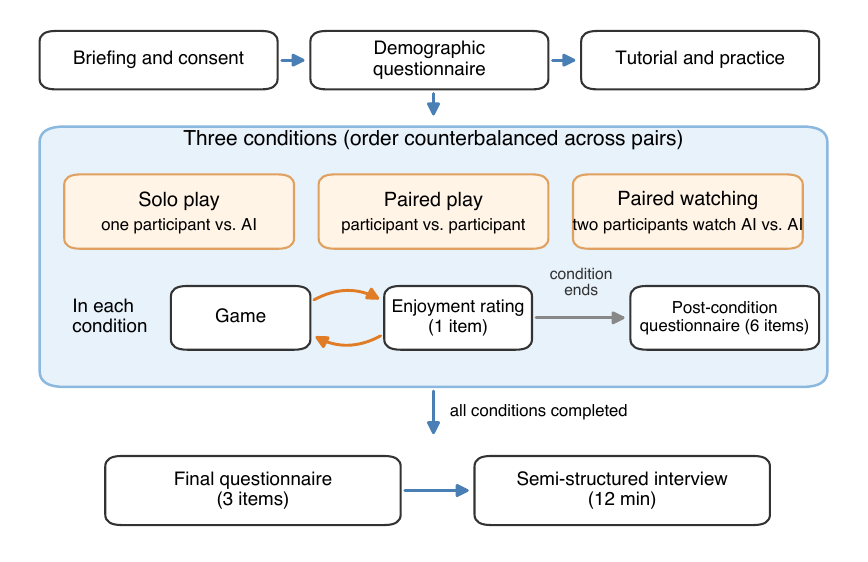}
\caption{Study procedure. The order of the three conditions was counterbalanced.}
\label{fig:study-procedure}
\end{figure}

Figure~\ref{fig:study-procedure} shows the procedure.
After a briefing and consent, a demographic questionnaire, a tutorial, and practice, participants completed the three conditions.
They rated enjoyment on one item after each game and answered the post-condition questionnaire (six items) after each condition.
After all three conditions, they answered the final questionnaire (three items) and took part in a 12-minute semi-structured interview.

\subsection{Measures}
Before the study, participants answered eight items on their demographics and their experience with games and generative AI (Table~\ref{tab:pre-post-items}).
Right after each game, we asked ``How enjoyable was the game you just played?'' on a 7-point scale from 1 (not enjoyable at all) to 7 (very enjoyable).
Participants rated before the changes for the next game were shown, and in the paired conditions they rated individually without seeing each other's answers.
Right after each condition, participants answered six 7-point items on their experience of speech and game changes (Table~\ref{tab:condition-items}).
After all conditions, they compared the three conditions and chose the one they would most like to replay, the one they would least like to replay, and the one in which their speech was best reflected (Table~\ref{tab:pre-post-items}).
The most and least preferred conditions had to be different.

\begin{table}[t]
\centering
\caption{Demographic questionnaire and final questionnaire.}
\label{tab:pre-post-items}
\small
\begin{tabular}{llp{0.6\linewidth}}
\toprule
Part & No. & Item \\
\midrule
Demographics & 1 & Age \\
& 2 & Gender \\
& 3 & Frequency of playing video games \\
& 4 & Experience with fighting games \\
& 5 & I think I am good at games (7-point) \\
& 6 & Frequency of using generative AI \\
& 7 & I think I can use generative AI well (7-point) \\
& 8 & Programming experience \\
\midrule
Final questionnaire & 1 & The condition I would most like to play again \\
& 2 & The condition I would least like to play again \\
& 3 & The condition in which my (our) speech was best reflected \\
\bottomrule
\end{tabular}
\end{table}

\begin{table}[t]
\centering
\caption{Items of the post-condition questionnaire (six items).}
\label{tab:condition-items}
\small
\begin{tabular}{llp{0.58\linewidth}}
\toprule
Construct & ID & Item \\
\midrule
Noticing changes & Q1 & In this condition, I noticed that the game changed. \\
Reflection of speech & Q2 & What I (we) said was reflected in the changes to the game. \\
Understanding of changes & Q3 & I understood why the game changed the way it did. \\
Match with intent & Q4 & The game changed in the direction I (we) wanted. \\
Sense of changing the game & Q5 & I felt that I (we) changed the game through what I (we) said. \\
Saying enough & Q6 & In this condition, I was able to say enough about what I thought of the game. \\
\bottomrule
\end{tabular}

\vspace{2pt}
\begin{minipage}{0.94\linewidth}\footnotesize
Endpoint labels were tailored to each item. 1 is the negative end and 7 the positive end.
\end{minipage}
\end{table}

After all conditions, we conducted a pair interview followed by individual interviews, and this paper analyzes the pair interviews (27 pairs).
The interviewer read out the pair's answers to the final questionnaire and asked why they had chosen the condition they most wanted to replay, the one they least wanted to replay, and the one in which their speech was best reflected.
Many pairs were then asked what the activity resembled, although the wording varied across pairs.
We also logged speech, LLM responses, applied setting changes, and per-game outcomes, durations, and remaining-health differences.
Because the paired conditions shared one microphone, speech was not attributed to individuals and was treated at the pair level.

\subsection{Analysis}
For all 54 participants, we performed the following two comparisons specified in the study plan.
\begin{align}
C_1 &= \text{Paired play} - \text{Solo play} \quad(\text{RQ1})\\
C_2 &= \text{Paired play} - \text{Paired watching} \quad(\text{RQ2})
\end{align}
These two contrasts are the primary analyses.
We report effect sizes, 95\% confidence intervals, and multiplicity-adjusted $p$-values.
Associations of wins and losses, speech volume, and change volume with subjective ratings are treated as additional exploratory analyses of the background of the experience.

Because post-game enjoyment is a 7-point ordinal scale, we used ordinal logistic mixed models.
Condition, game progress within the condition, their interaction, and order were fixed effects, and we accounted for correlations due to participants, pairs, and shared games.
Because the third game of a best-of-three match occurs only after a 1--1 split, the primary analysis used the first and second games of each match.
We exponentiated the log odds ratio averaged over the distribution of game progress in the primary analysis and report it as the condition difference across target games.
We applied the Holm correction to the seven tests on enjoyment (condition differences, progress, and interactions).
For the paired conditions, we computed the mean and the absolute difference of the two ratings for each game and averaged them over the six games of each pair.
We describe the means, standard deviations (SD), and distributions across the 27 pairs.

The post-condition questionnaire (six items) was analyzed item by item with ordinal logistic mixed models.
Condition and order were fixed effects, with random intercepts for pairs and participants within pairs, and we applied the Holm correction to the 12 tests formed by six items and two contrasts.
The reported 95\% confidence intervals are unadjusted.
We checked sensitivity to the analysis method with models treating ratings as continuous, pair-level resampling, an additional random effect for the shared experience in paired conditions, and the addition of third games for enjoyment.

The final questionnaire asked participants to choose one of three conditions, so we tallied the number and proportion of choices for each condition.
Speech and change volume were summarized as characters of speech per minute and changes per game.
For solo play, we averaged the rates of the two individual sessions in a pair.
For the paired conditions, we used the rate of the joint session.

As an exploration of the background of enjoyment, we analyzed associations of ratings with wins and losses, game duration, and remaining-health difference normalized by maximum health.
We separated each participant--condition mean from per-game deviations from that mean and estimated within-participant associations by condition with ordinal mixed models.
Accounting for order, game progress, and participants, pairs, shared games, and shared experience in paired conditions, we applied the Holm correction to the nine associations of three indicators and three conditions.
In addition, using 81 observations from 27 pairs and three conditions, we estimated six associations of speech and change volume with mean post-game enjoyment, ``I was able to say enough about what I thought of the game'' (Q6), and ``What I (we) said was reflected in the changes to the game'' (Q2), with linear models adjusting for pair, condition, and order.
Speech volume was transformed as $\log_2(1+x)$, and the two predictors were entered in separate models.
For the slopes, we computed individual 95\% intervals from 4,000 pair-level bootstrap resamples.

The interviews were transcribed and the reasons given by each pair were organized according to the three stages described in the introduction.
Tallies were made at the pair level, and frequencies are reported as the number of pairs that mentioned each reason.

\section{Results}
All 27 pairs (54 participants) were included in the analysis.
There were 162 post-condition questionnaires with no missing data.
The primary analysis of post-game enjoyment used 972 ratings from the first and second games of each match.

\subsection{Enjoyment of Play and Replay Preference}
\begin{figure*}[t]
\centering
\includegraphics[width=\textwidth]{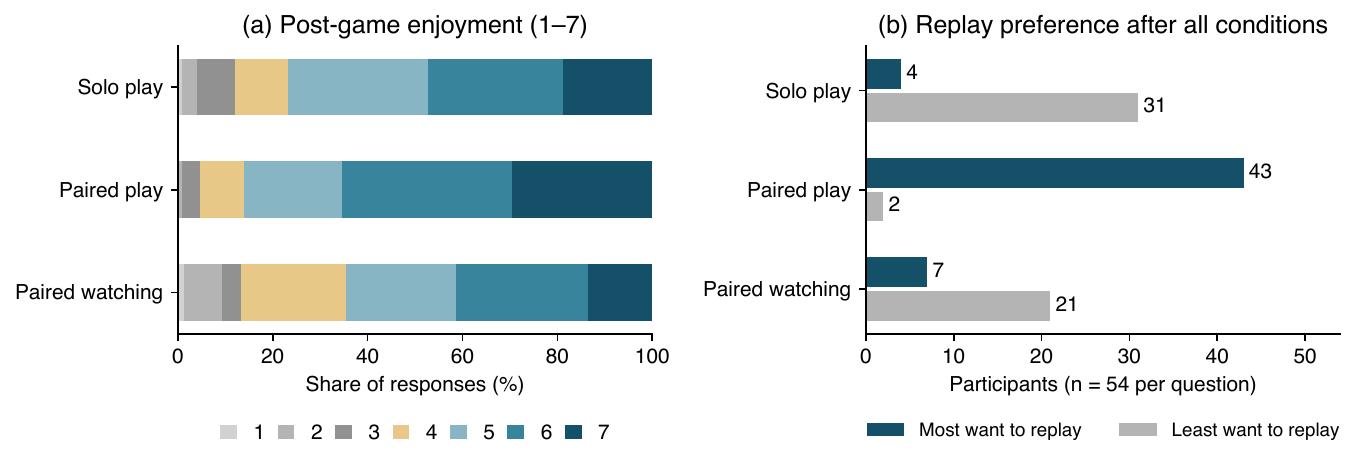}
\caption{Enjoyment of play and replay preference. (a) Distribution of 324 ratings per condition from the first and second games, including repeated ratings from the same 54 participants. (b) Number of participants who chose each condition in the final questionnaire.}
\label{fig:enjoyment-replay}
\end{figure*}

Mean enjoyment was 5.26 in solo play, 5.75 in paired play, and 4.96 in paired watching (Figure~\ref{fig:enjoyment-replay}a).
The condition difference across target games was $OR=3.18$ (95\% CI $[2.06, 4.90]$) for paired versus solo play and $OR=6.68$ (95\% CI $[4.10, 10.87]$) for paired play versus paired watching.
Enjoyment was higher in paired play in both contrasts (Holm-adjusted $p<.001$ for both).
These results concern the enjoyment of play as rated right after each game.

In the final questionnaire, 43 of 54 participants (79.6\%) chose paired play as the condition they would most like to replay.
Only two chose paired play as the condition they would least like to replay (Figure~\ref{fig:enjoyment-replay}b).

\subsection{Experiences of Speech and Changes}
Table~\ref{tab:condition-results} shows all six items of the post-condition questionnaire.
Compared with solo play, paired play received higher ratings for ``I was able to say enough about what I thought of the game'' (Q6; $OR=4.96$, 95\% CI $[2.33, 10.54]$) and ``What I (we) said was reflected in the changes to the game'' (Q2; $OR=3.09$, 95\% CI $[1.47, 6.50]$), and these differences remained after Holm correction across the 12 tests.
No clear differences were found for the remaining four items or for any of the six items between paired play and paired watching.

\begin{table*}[t]
\centering\small
\caption{Condition differences in the post-condition questionnaire (six items). $C_1$: paired play $-$ solo play; $C_2$: paired play $-$ paired watching.}
\label{tab:condition-results}
\begin{tabular}{lrrrr}
\toprule
Item & $C_1$: OR [95\% CI] & Adjusted $p$ & $C_2$: OR [95\% CI] & Adjusted $p$ \\
\midrule
Q1 Noticing changes$^{\dagger}$ & 2.20 [1.04, 4.68] & 0.319 & 0.65 [0.29, 1.43] & 1.000 \\
Q2 Reflection of speech & 3.09 [1.47, 6.50] & 0.033 & 1.33 [0.65, 2.69] & 1.000 \\
Q3 Understanding of changes & 2.23 [1.07, 4.67] & 0.294 & 1.45 [0.71, 2.98] & 1.000 \\
Q4 Match with intent & 1.29 [0.65, 2.57] & 1.000 & 0.76 [0.38, 1.52] & 1.000 \\
Q5 Sense of changing the game$^{\dagger}$ & 2.88 [1.33, 6.20] & 0.071 & 0.98 [0.46, 2.10] & 1.000 \\
Q6 Saying enough & 4.96 [2.33, 10.54] & $<.001$ & 1.07 [0.53, 2.18] & 1.000 \\
\bottomrule
\end{tabular}
\vspace{2pt}
\begin{minipage}{0.94\textwidth}\footnotesize
An OR greater than 1 means higher ratings are more likely in paired play. Models adjust for order and include random intercepts for pairs and participants within pairs.
$p$-values are Holm-corrected over the 12 tests (6 items $\times$ 2 contrasts). The 95\% CIs are unadjusted.
$^{\dagger}$ Approximate diagnostics with independent models raised concerns about the proportional odds assumption, so a single OR should be interpreted with caution.
\end{minipage}
\end{table*}

In a sensitivity analysis adding a random effect for shared experience, the intervals for enjoyment widened, but the direction of the two condition differences and of the two item differences above was maintained.
However, when Q2 ratings were treated as continuous, the paired- versus solo-play difference had a Holm-adjusted $p=.063$, so the certainty of this difference depended on the analysis method.

Characters of speech per minute were 58 in solo play, 130 in paired play, and 142 in paired watching, and changes per game were 3.19, 4.33, and 4.33, respectively.
Speech volume is the amount for the whole session.
We did not separate the amount per person.
The number of changes is the amount of applied setting changes.
In the exploratory analysis, more speech was associated with higher ratings of ``I was able to say enough about what I thought of the game'' (Q6).
Associations of speech and change volume with post-game enjoyment and with ``What I (we) said was reflected in the changes to the game'' (Q2) were unclear.

\subsection{Dyadic Enjoyment, Winning, and Losing}
\begin{figure}[t]
\centering
\includegraphics[width=0.9\linewidth]{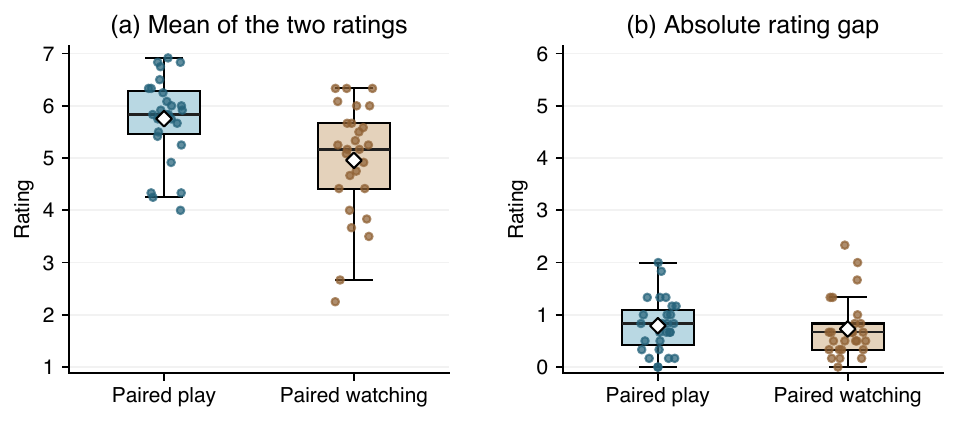}
\caption{(a) Mean and (b) absolute difference of the two players' enjoyment ratings. Values were computed from the two ratings in each game and averaged over each pair's six games (first and second games of each match). Each dot is a pair (27 pairs per condition), jittered horizontally. Boxes show the interquartile range, horizontal lines the median, whiskers the most extreme values within 1.5 times the interquartile range from the box, and white diamonds the mean. Mean (SD): (a) paired play 5.75 (0.81), paired watching 4.96 (1.08); (b) paired play 0.79 (0.52), paired watching 0.73 (0.56). SDs are across the 27 pairs.}
\label{fig:shared-enjoyment}
\end{figure}

The mean (SD) of the two players' enjoyment, averaged over each pair's six games, was 5.75 (0.81) in paired play and 4.96 (1.08) in paired watching (Figure~\ref{fig:shared-enjoyment}a).
Averaging the absolute difference between the two ratings of the same game for each pair gave 0.79 (SD 0.52) in paired play and 0.73 (SD 0.56) in paired watching (Figure~\ref{fig:shared-enjoyment}b).
The SDs represent variation across the 27 pairs.

In the exploratory analysis of wins and losses, comparing wins and losses of the same participant gave $OR=5.40$ (95\% CI $[3.00, 9.74]$) in solo play and $OR=3.96$ (95\% CI $[2.26, 6.95]$) in paired play (Holm-adjusted $p<.001$ for both).
This analysis captures the association between winning and enjoyment within each condition.
We did not test how much winning and losing explain the differences in enjoyment between conditions.

\subsection{Pair Interviews}
In the interviews, the interviewer read out the pair's answers to the final questionnaire and asked why they had chosen the condition they most wanted to replay, the one they least wanted to replay, and the one in which their speech felt best reflected.
Following the three stages, we describe (1) talking, (2) requests taking shape through exchanges with the partner, (3) cues for feeling that speech was reflected, and (4) the purpose of changes and how the activity was understood.
Counts are the number of pairs that mentioned each point, and a pair may mention more than one.
Points (1), (2), and (4) were raised spontaneously by participants and were not asked of every pair.
Quotes are labeled with pair numbers (G1--G27) and translated from Japanese.

\begin{table}[t]
\centering
\caption{Reasons given in the interviews and the number of pairs (out of 27) that mentioned them. A pair may mention more than one reason.}
\label{tab:interview-codes}
\small
\begin{tabular}{p{0.75\linewidth}r}
\toprule
Reason & Pairs \\
\midrule
\multicolumn{2}{l}{\textbf{Why participants wanted to replay paired play}} \\
Being able to talk with the partner; getting reactions back & 15 \\
Back-and-forth of changing the game together & 9 \\
Being able to control the character and check what changed & 9 \\
Enjoying playing against a person rather than a CPU & 8 \\
\midrule
\multicolumn{2}{l}{\textbf{Why participants did not want to replay solo play}} \\
Hard to keep talking alone; not used to talking to oneself & 10 \\
No partner; no stimulation & 5 \\
\midrule
\multicolumn{2}{l}{\textbf{Why participants did not want to replay paired watching}} \\
Unable to control the character; frustrating & 12 \\
Unable to check the changes oneself & 6 \\
\midrule
\multicolumn{2}{l}{\textbf{Cues for feeling that speech was best reflected}} \\
Being able to check or feel the changes through one's own play & 10 \\
Having room to watch; seeing the whole match & 9 \\
Having talked a lot; being able to express requests & 9 \\
Clearly visible changes such as appearance or new moves & 7 \\
\bottomrule
\end{tabular}
\end{table}

\paragraph{Talking.}
The most common reason for wanting to replay paired play was being able to talk with the partner (15 pairs; Table~\ref{tab:interview-codes}).
This was often contrasted with solo play, and the most common reason for not wanting to replay solo play was the difficulty of keeping talking alone (10 pairs).
One pair said that alone ``you focus on your own play, so you don't talk much,'' whereas as a pair ``you talk naturally'' (G13).
Others said that alone they did not know what to say or could not come up with ideas (7 pairs), and one pair said that ``while talking, my own thoughts come together'' (G25).
For paired watching, some said it was easier to talk because they ``could just concentrate on watching'' without being absorbed in play (8 pairs; e.g., G10), whereas others said that nothing came to mind when only watching or that their talk became closer to heckling (G16, G17, G19).
Some participants said they could talk enough even alone (G10, G18, G24), so the difficulty of talking varied across participants.

\paragraph{Requests taking shape through exchanges with the partner.}
In the paired conditions, participants described the partner's remarks as helping their own requests take shape.
One pair said that even when one of them ``said something vague and abstract,'' the partner turned it into a concrete form (G18).
In another pair, the partner's way of phrasing served as ``a model'' (G20).
Another pair said that without a partner they were ``not stimulated'' and requests did not come out (G21).
One pair enjoyed the ``back-and-forth'' of countering each other's countermeasures (G18), and another thought about their next strategy while listening to the partner's requests (G26).
On the other hand, some described each wanting to strengthen the character they controlled (G10), or noted that paired play could become ``whoever says it first wins'' (G15).
Exchanges with the partner did not always converge on a single request.

\paragraph{Cues for feeling that speech was reflected.}
In the final questionnaire, 26 participants chose paired play, 17 paired watching, and 11 solo play as the condition in which their speech was best reflected.
The cues participants gave for this choice differed by condition.
For playing, they mentioned being able to check the changes through their own play (10 pairs), for example being able to ``try out exactly what we said with our own hands'' (G19).
For watching, they mentioned being able to see the whole match without being absorbed in play (9 pairs).
However, in watching, a changed move could not be checked unless the AI used it, and one participant said, ``we changed it but never got to see it'' (G19; similar remarks in 5 pairs).
Other cues were having talked a lot (9 pairs) and visible changes such as appearance or new moves (7 pairs), while fine numerical changes such as frame-level speed were said to be hard to notice (G24).
The reasons for not wanting to replay paired watching were likewise being unable to control the character (12 pairs) and being unable to check the changes oneself (6 pairs).

\paragraph{Purpose of changes and how the activity was understood.}
Several pairs said that in solo play, changes went in the direction of strengthening themselves (7 pairs).
One pair said that in solo play they thought ``just about how to win,'' but in paired play they thought about ``how the game could become more fun'' (G7).
In paired play, one pair said that because unreasonable changes come back to oneself, ``we'll suffer the unfairness too, so we want to get the balance right'' (G22).
The same pair said that in solo play only they became stronger, and ``that's not fun either.''
On the other hand, some pairs thought about winning even in paired play, or one player alone became stronger (4 pairs).
One participant felt ``sorry'' about being the only one who became stronger but still enjoyed it (G9), and others mentioned that changes closing the skill gap made the game enjoyable (3 pairs).

How the activity was understood reflected the form of engagement.
Watching was described as being a coach, a director, or someone raising characters, whereas playing was described as being a player (4 pairs).
One pair said that ``a player and a director are probably different,'' explaining that a director adjusts the overall balance while a player tries to sharpen their own strengths (G8).
In the paired conditions, some said that the one they were making the game with was not the AI but the participant next to them (3 pairs; e.g., G10, G25).

\section{Discussion}

\subsection{Why Paired Play Was Preferred}
Regarding RQ1, paired play received higher ratings than solo play for post-game enjoyment and for ``I was able to say enough about what I thought of the game'' (Q6).
Ratings of ``What I (we) said was reflected in the changes to the game'' (Q2) also tended to be higher, but the strength of the evidence depended on the analysis method, and the other four items showed no clear differences.
Thus, the benefits of joint participation do not necessarily extend to the experience of changes as a whole.
This gap between enjoyment and the questionnaire indicates the need to treat the enjoyment of play and the experience of changing the game by stating one's intentions as distinct.

Paired play was also the condition most often chosen for replay.
Playing against a person, winning and losing, conversation with the partner, shared events, and co-editing may all contribute to this.
In solo play the opponent was a CPU, so our comparison captures differences in the overall participation experience that combine these factors.
In the interviews, the reasons for choosing paired play overlapped---being able to talk with the partner, the back-and-forth of changing the game together, being able to check changes through one's own play, and the enjoyment of playing against a person---and could not be reduced to a single factor.
Few pairs cited winning or losing as a reason.
The main reason for avoiding solo play was the difficulty of continuing to talk alone, which is consistent with the lower speech volume and lower ratings of ``I was able to say enough about what I thought of the game'' (Q6) in solo play.

Regarding RQ2, enjoyment in paired watching was lower than in paired play even though speech volume was highest in paired watching, and the post-condition questionnaire showed no clear differences.
In the interviews, the reasons for avoiding watching were being unable to control the character and being unable to check the changes oneself.
Watching allows participants to concentrate on talking, but offers no opportunity to check the outcome of changes through their own play.
Moreover, the cues for feeling that speech was reflected differed by condition.
When playing, participants checked changes through play, and when watching, they looked at the whole match.
That the post-condition questionnaire showed no clear differences between paired play and paired watching may therefore indicate not that the two conditions provided similar experiences, but that participants gave similar ratings on different grounds.
Watching was described as being a ``coach'' or a ``director,'' and playing as being a ``player.''
The difference in the form of engagement was experienced as a difference in one's position in changing the game.

\subsection{Winning, Losing, and Edits Both Players Can Enjoy}
The association between winning and higher enjoyment suggests that post-game ratings include reactions to the match outcome.
This analysis, however, does not explain the overall condition differences by winning and losing.
The mean of the two players' ratings was higher in paired play, whereas the gap between the two ratings was similar in both conditions.
Because the gap was tallied descriptively, this does not establish equivalence, but the overall level of a pair's ratings and the gap between the two players need to be considered separately.
In co-editing, a design challenge is how players can discuss changes that both can enjoy, beyond changes that make one side more likely to win.
In the interviews, several pairs said that in solo play, changes went in the direction of strengthening themselves.
In paired play, participants said they considered directions both could enjoy, because changes come back to oneself and the partner's reactions are visible.
At the same time, some pairs thought about winning even in paired play, or one player alone became stronger, so edits that both enjoy do not necessarily arise on their own.
Some participants mentioned that changes closing the skill gap made the game enjoyable, suggesting that adjusting imbalances in winning and losing can be one purpose of editing.

In the exploratory analysis, speech volume was associated only with ``I was able to say enough about what I thought of the game'' (Q6), and neither speech volume nor the number of changes was clearly associated with post-game enjoyment or with ``What I (we) said was reflected in the changes to the game'' (Q2).
Thus, there is insufficient basis for making the increase of speech or changes alone a design goal for enjoyable editing.
It is necessary to examine whether intentions were shared, how the system understood them, and whether the changes were accepted.
In the interviews, participants said that the partner's remarks served as cues or models that made their requests concrete.
This is consistent with our framing of requests that take shape in a conversation between two players.
However, some described each wanting to strengthen their own character, or that paired play could become ``whoever says it first wins,'' so the two players' requests did not always converge.
How the two players' requests were adopted and how disagreements were handled were not sufficiently captured by the interviews.

\section{Limitations and Future Work}

\subsection{Limitations}
This study involved 27 pairs (54 participants) of acquaintances, so the estimates are uncertain and the scope of application is constrained by the relationships between participants.
Between solo and paired play, the number of people, the opponent, and the shared experience change simultaneously.
Separating the effects of these factors requires additional conditions.

Multiplicity adjustment was limited to within each family of tests, and the 95\% intervals are individual intervals.
The post-condition questionnaire consists of single-item measures developed for this study, and some items raise concerns about the proportional odds assumption.
In addition, the presence of a body-size handicap that was removed partway through the study coincides with the period of data collection.
Although sensitivity analyses confirmed the direction of the results, the model assumptions and the influence of the data-collection period warrant caution.

Enjoyment was measured for the immediately preceding game, and our design cannot separate the enjoyment of editing from that of play.
Because speech in the paired conditions was recorded jointly, attributing speech to individuals is limited.
The associations of winning, speech, and changes with ratings are observational, and confounding or reverse relationships remain possible.
Testing how editing increases enjoyment requires measurements that link the content of changes to participants' intentions and evaluations.

The interviews were conducted in pairs, so answers may have been shaped by consideration for the partner.
The interviews focused on the reasons for participants' choices and did not directly ask about disagreements between the two players or how they perceived winning and losing.
Our descriptions of these topics are limited to what participants raised spontaneously.
The interview transcripts may contain recognition errors and speaker misattributions.

\subsection{Future Work}
Future studies should use a design that measures editing and play separately and capture the experiences of both winners and losers.
Recording pre-edit intentions, ratings of understanding and reflection, acceptance, rejection, or revision of changes, and play after changes in a linked manner would allow us to examine which editing processes lead to satisfaction.
How the two players' requests were adopted and how disagreements were handled, which the interviews did not sufficiently capture, should be examined by linking speech, the LLM's rationales for changes, and the actual changes.
Because participants said it was difficult to check the outcome of changes when watching, ways of presenting changes so that they can be verified also need to be explored.
When conducting studies in public settings, the number of people, the relationships between participants, and the roles in and opportunities for playing and editing should also be recorded.

It is also necessary to evaluate what kinds of games can be made.
The number of changes and the categories of settings capture the amount and scope of editing activity.
To evaluate the resulting games, we could use whether play holds together, originality, ease of understanding the rules, and enjoyment for others.
Saving the final settings and games and adding evaluations by people who did not take part in making them remain future work.

\section{Conclusion}
Through a study with 27 pairs (54 participants) using PATCH FIGHTER, we compared post-game enjoyment and experiences of speech and changes in solo play, paired play, and paired watching.
Paired play yielded higher post-game enjoyment than solo play and paired watching and was the condition most often chosen for replay.
It also received higher ratings than solo play for ``I was able to say enough about what I thought of the game'' (Q6).
However, because our design cannot separate the enjoyment of editing from that of play, the trend in enjoyment concerns the overall experience of playing while changing the game.
Ratings also differed between wins and losses, indicating the need for a design that measures editing and play separately.
Pair interviews suggested that paired play was preferred for the combination of talking with the partner, changing the game together, and checking changes through one's own play, whereas watching left participants unable to check the changes themselves.
In future work, we will use such a design, link speech, the LLM's rationales, and the applied changes, and examine through public studies the process of co-editing and the enjoyment and diversity of the games created.

\bibliographystyle{unsrtnat}
\bibliography{references}

\begin{thebibliography}{27}
\providecommand{\natexlab}[1]{#1}
\providecommand{\url}[1]{\texttt{#1}}
\expandafter\ifx\csname urlstyle\endcsname\relax
  \providecommand{\doi}[1]{doi: #1}\else
  \providecommand{\doi}{doi: \begingroup \urlstyle{rm}\Url}\fi

\bibitem[Gallotta et~al.(2024{\natexlab{a}})Gallotta, Todd, Zammit, Earle,
  Liapis, Togelius, and Yannakakis]{gallotta2024llmgames}
Roberto Gallotta, Graham Todd, Marvin Zammit, Sam Earle, Antonios Liapis,
  Julian Togelius, and Georgios~N. Yannakakis.
\newblock Large language models and games: A survey and roadmap.
\newblock \emph{IEEE Transactions on Games}, 2024{\natexlab{a}}.
\newblock \doi{10.1109/TG.2024.3461510}.

\bibitem[Sudhakaran et~al.(2023)Sudhakaran, Gonz{\'a}lez-Duque, Freiberger,
  Glanois, Najarro, and Risi]{sudhakaran2023mariogpt}
Shyam Sudhakaran, Miguel Gonz{\'a}lez-Duque, Matthias Freiberger, Claire
  Glanois, Elias Najarro, and Sebastian Risi.
\newblock {MarioGPT}: Open-ended text2level generation through large language
  models.
\newblock In \emph{Advances in Neural Information Processing Systems},
  volume~36, pages 54213--54227. Curran Associates, Inc., 2023.
\newblock \doi{10.52202/075280-2361}.

\bibitem[Gallotta et~al.(2024{\natexlab{b}})Gallotta, Liapis, and
  Yannakakis]{gallotta2024llmaker}
Roberto Gallotta, Antonios Liapis, and Georgios~N. Yannakakis.
\newblock {LLMaker}: A game level design interface using (only) natural
  language.
\newblock In \emph{2024 IEEE Conference on Games (CoG)}, pages 1--2,
  2024{\natexlab{b}}.
\newblock \doi{10.1109/CoG60054.2024.10645626}.

\bibitem[Whitehead et~al.(2025)]{whitehead2025conversational}
Jim Whitehead et~al.
\newblock Conversational interactions with procedural generators using large
  language models.
\newblock In \emph{Proceedings of the 20th International Conference on the
  Foundations of Digital Games (FDG '25)}, pages 1--7. ACM, 2025.
\newblock \doi{10.1145/3723498.3723788}.

\bibitem[Rezwana and Maher(2023)]{rezwana2023cofi}
Jeba Rezwana and Mary~Lou Maher.
\newblock Designing creative {AI} partners with {COFI}: A framework for
  modeling interaction in human-{AI} co-creative systems.
\newblock \emph{ACM Transactions on Computer-Human Interaction}, 30\penalty0
  (5):\penalty0 1--28, 2023.
\newblock \doi{10.1145/3519026}.

\bibitem[Draxler et~al.(2024)Draxler, Werner, Lehmann, Hoppe, Schmidt, Buschek,
  and Welsch]{draxler2024ghostwriter}
Fiona Draxler, Anna Werner, Florian Lehmann, Matthias Hoppe, Albrecht Schmidt,
  Daniel Buschek, and Robin Welsch.
\newblock The {AI} ghostwriter effect: When users do not perceive ownership of
  {AI}-generated text but self-declare as authors.
\newblock \emph{ACM Transactions on Computer-Human Interaction}, 31\penalty0
  (2):\penalty0 1--40, 2024.
\newblock \doi{10.1145/3637875}.

\bibitem[Torii et~al.(2023)Torii, Murakami, and Ochiai]{torii2023lottery}
Maya~Grace Torii, Takahito Murakami, and Yoichi Ochiai.
\newblock Lottery and sprint: Generate a board game with design sprint method
  on {AutoGPT}.
\newblock In \emph{Companion Proceedings of the Annual Symposium on
  Computer-Human Interaction in Play (CHI PLAY Companion '23)}, pages 259--265.
  ACM, 2023.
\newblock \doi{10.1145/3573382.3623706}.

\bibitem[Torii et~al.(2026)Torii, Murakami, and Ochiai]{torii2026arcade}
Maya~Grace Torii, Takahito Murakami, and Yoichi Ochiai.
\newblock Lottery and sprint arcade: Enabling player-driven game editing with
  generative {AI}.
\newblock \emph{The Journal of the Society for Art and Science}, 25\penalty0
  (2):\penalty0 12:1--12:14, 2026.
\newblock \doi{10.3756/artsci.25.12_1}.

\bibitem[Yannakakis and Togelius(2011)]{yannakakis2011edpcg}
Georgios~N. Yannakakis and Julian Togelius.
\newblock Experience-driven procedural content generation.
\newblock \emph{IEEE Transactions on Affective Computing}, 2\penalty0
  (3):\penalty0 147--161, 2011.
\newblock \doi{10.1109/T-AFFC.2011.6}.

\bibitem[Jennings et~al.(2024)Jennings, Wang, Li, Smith, and
  Hartmann]{jennings2024whatsthegame}
Nicholas Jennings, Han Wang, Isabel Li, James Smith, and Bjoern Hartmann.
\newblock What's the game, then? {Opportunities} and challenges for runtime
  behavior generation.
\newblock In \emph{Proceedings of the 37th Annual ACM Symposium on User
  Interface Software and Technology (UIST '24)}, pages 1--13. ACM, 2024.
\newblock \doi{10.1145/3654777.3676358}.

\bibitem[Liebers et~al.(2024)Liebers, Pf{\"u}tzenreuter, Auda, Gruenefeld, and
  Schneegass]{liebers2024computer}
Carina Liebers, Niklas Pf{\"u}tzenreuter, Jonas Auda, Uwe Gruenefeld, and
  Stefan Schneegass.
\newblock ``{Computer}, generate!'' -- investigating user-controlled generation
  of immersive virtual environments.
\newblock In \emph{HHAI 2024: Hybrid Human AI Systems for the Social Good},
  Frontiers in Artificial Intelligence and Applications. IOS Press, 2024.
\newblock \doi{10.3233/FAIA240196}.

\bibitem[van Rozen(2025)]{vanrozen2025live}
Riemer van Rozen.
\newblock Live game design: Prototyping at the speed of play.
\newblock In \emph{Proceedings of the 20th International Conference on the
  Foundations of Digital Games (FDG '25)}, pages 1--12. ACM, 2025.
\newblock \doi{10.1145/3723498.3723726}.

\bibitem[Chen et~al.(2025)Chen, Zhang, Wang, Huo, Li, and
  Wang]{chen2023gamegpt}
Dake Chen, Haoyang Zhang, Hanbin Wang, Yunhao Huo, Yuzhao Li, and Junjie Wang.
\newblock {GameGPT}: Multi-agent collaborative framework for game development,
  2025.
\newblock arXiv:2310.08067v2.

\bibitem[Hong et~al.(2025)Hong, Wu, and Zhao]{hong2025chatge}
Jiale Hong, Hongqiu Wu, and Hai Zhao.
\newblock Game development as human-{LLM} interaction.
\newblock In \emph{Proceedings of the 63rd Annual Meeting of the Association
  for Computational Linguistics (Volume 1: Long Papers)}, pages 4333--4354,
  2025.
\newblock \doi{10.18653/v1/2025.acl-long.218}.

\bibitem[Huang et~al.(2026)Huang, Li, Li, Wang, Chen, Ge, Si, Shen, Yang, Wang,
  and Guo]{huang2026play2code}
Yixu Huang, Bo~Li, Na~Li, Zhe Wang, Kaijie Chen, Haonan Ge, Qingyi Si, Yuanzhe
  Shen, Ruihan Yang, Guangjing Wang, and Hongcheng Guo.
\newblock {GUI} agents for continual game generation, 2026.
\newblock arXiv:2605.28258.

\bibitem[Wu et~al.(2026)Wu, Fu, You, Zhou, Liu, Salvi, Lin, Zhang, Lan, Zhu,
  Zhong, She, and Huang]{wu2026rsigame}
Wenyi Wu, Minghao Fu, Jieyu You, Kun Zhou, Siqi Liu, Aayush Salvi, Yiheng Lin,
  Ce~Zhang, Xiaohan Lan, Jiahui Zhu, Yujie Zhong, Qi~She, and Biwei Huang.
\newblock {RSIGame}: Autonomous agentic game development with recursive
  self-improvement, 2026.
\newblock arXiv:2609.39045.

\bibitem[Liu et~al.(2023)Liu, Sarkar, Negreanu, Zorn, Williams, Toronto, and
  Gordon]{liu2023wants}
Michael~Xieyang Liu, Advait Sarkar, Carina Negreanu, Benjamin Zorn, Jack
  Williams, Neil Toronto, and Andrew~D. Gordon.
\newblock ``{What} it wants me to say'': Bridging the abstraction gap between
  end-user programmers and code-generating large language models.
\newblock In \emph{Proceedings of the 2023 CHI Conference on Human Factors in
  Computing Systems (CHI '23)}, pages 1--31. ACM, 2023.
\newblock \doi{10.1145/3544548.3580817}.

\bibitem[Zamfirescu-Pereira et~al.(2023)Zamfirescu-Pereira, Wong, Hartmann, and
  Yang]{zamfirescu2023johnny}
J.D. Zamfirescu-Pereira, Richmond~Y. Wong, Bjoern Hartmann, and Qian Yang.
\newblock Why {Johnny} can't prompt: How non-{AI} experts try (and fail) to
  design {LLM} prompts.
\newblock In \emph{Proceedings of the 2023 CHI Conference on Human Factors in
  Computing Systems (CHI '23)}, pages 1--21. ACM, 2023.
\newblock \doi{10.1145/3544548.3581388}.

\bibitem[Subramonyam et~al.(2024)Subramonyam, Pea, Pondoc, Agrawala, and
  Seifert]{subramonyam2024gulf}
Hari Subramonyam, Roy Pea, Christopher Pondoc, Maneesh Agrawala, and Colleen
  Seifert.
\newblock Bridging the gulf of envisioning: Cognitive challenges in prompt
  based interactions with {LLMs}.
\newblock In \emph{Proceedings of the CHI Conference on Human Factors in
  Computing Systems (CHI '24)}, pages 1--19. ACM, 2024.
\newblock \doi{10.1145/3613904.3642754}.

\bibitem[Clark and Brennan(1991)]{clark1991grounding}
Herbert~H. Clark and Susan~E. Brennan.
\newblock Grounding in communication.
\newblock In Lauren~B. Resnick, John~M. Levine, and Stephanie~D. Teasley,
  editors, \emph{Perspectives on Socially Shared Cognition}, pages 127--149.
  American Psychological Association, 1991.
\newblock \doi{10.1037/10096-006}.

\bibitem[Miyake(1986)]{miyake1986constructive}
Naomi Miyake.
\newblock Constructive interaction and the iterative process of understanding.
\newblock \emph{Cognitive Science}, 10\penalty0 (2):\penalty0 151--177, 1986.
\newblock \doi{10.1207/s15516709cog1002_2}.

\bibitem[Han et~al.(2024)Han, Qiu, Cheng, and LC]{han2024teams}
Yuanning Han, Ziyi Qiu, Jiale Cheng, and RAY LC.
\newblock When teams embrace {AI}: Human collaboration strategies in generative
  prompting in a creative design task.
\newblock In \emph{Proceedings of the CHI Conference on Human Factors in
  Computing Systems (CHI '24)}, pages 1--14. ACM, 2024.
\newblock \doi{10.1145/3613904.3642133}.

\bibitem[Zhu et~al.(2025)Zhu, Osgood, and Callison-Burch]{zhu2025overhearing}
Andrew Zhu, Evan Osgood, and Chris Callison-Burch.
\newblock First steps towards overhearing {LLM} agents: A case study with
  {Dungeons \& Dragons} gameplay, 2025.
\newblock arXiv:2505.22809.

\bibitem[Gajadhar et~al.(2008)Gajadhar, de~Kort, and
  IJsselsteijn]{gajadhar2008shared}
Brian~J. Gajadhar, Yvonne A.~W. de~Kort, and Wijnand~A. IJsselsteijn.
\newblock Shared fun is doubled fun: Player enjoyment as a function of social
  setting.
\newblock In \emph{Fun and Games 2008}, Lecture Notes in Computer Science,
  pages 106--117. Springer, 2008.
\newblock \doi{10.1007/978-3-540-88322-7_11}.

\bibitem[Reeves et~al.(2005)Reeves, Benford, O'Malley, and
  Fraser]{reeves2005spectator}
Stuart Reeves, Steve Benford, Claire O'Malley, and Mike Fraser.
\newblock Designing the spectator experience.
\newblock In \emph{Proceedings of the SIGCHI Conference on Human Factors in
  Computing Systems (CHI '05)}, pages 741--750. ACM, 2005.
\newblock \doi{10.1145/1054972.1055074}.

\bibitem[Cheung and Huang(2011)]{cheung2011starcraft}
Gifford Cheung and Jeff Huang.
\newblock Starcraft from the stands: Understanding the game spectator.
\newblock In \emph{Proceedings of the SIGCHI Conference on Human Factors in
  Computing Systems (CHI '11)}, pages 763--772. ACM, 2011.
\newblock \doi{10.1145/1978942.1979053}.

\bibitem[Paliyawan et~al.(2024)Paliyawan, Thawonmas, Sookhanaphibarn, and
  Choensawat]{paliyawan2024apg}
Pujana Paliyawan, Ruck Thawonmas, Kingkarn Sookhanaphibarn, and Worawat
  Choensawat.
\newblock Audience participation fighting game: Exploring social facilitation
  for an enhanced {APG} experience.
\newblock \emph{Heliyon}, 10\penalty0 (2):\penalty0 e23967, 2024.
\newblock \doi{10.1016/j.heliyon.2023.e23967}.

\end{thebibliography}

\end{document}